\documentclass[twocolumn,preprintnumbers,amssymb,amsmath,9pt,superscriptaddress]{revtex4}[10pt]
\usepackage{graphicx,psfrag}
\usepackage{dcolumn}
\usepackage{bm}
\begin{document} 
\input epsf.tex
\newcommand{\beq}{\begin{eqnarray}}
\newcommand{\eeq}{\end{eqnarray}}
\newcommand{\nn}{\nonumber}
\def\ltap{\ \raise.3ex\hbox{$<$\kern-.75em\lower1ex\hbox{$\sim$}}\ }
\def\gtap{\ \raise.3ex\hbox{$>$\kern-.75em\lower1ex\hbox{$\sim$}}\ }
\def\CO{{\cal O}}
\def\CL{{\cal L}}
\def\CM{{\cal M}}
\def\tr{{\rm\ Tr}}
\def\CO{{\cal O}}
\def\CL{{\cal L}}
\def\CM{{\cal M}}
\def\mpl{M_{\rm Pl}}
\newcommand{\bel}[1]{\be\label{#1}}
\def\al{\alpha}
\def\bt{\beta}
\def\eps{\epsilon}
\def\eg{{\it e.g.}}
\def\ie{{\it i.e.}}
\def\mn{{\mu\nu}}
\newcommand{\rep}[1]{{\bf #1}}
\def\be{\begin{equation}}
\def\ee{\end{equation}}
\def\bea{\begin{eqnarray}}
\def\eea{\end{eqnarray}}
\newcommand{\eref}[1]{(\ref{#1})}
\newcommand{\Eref}[1]{Eq.~(\ref{#1})}
\newcommand{\gsim}{ \mathop{}_{\textstyle \sim}^{\textstyle >} }
\newcommand{\lsim}{ \mathop{}_{\textstyle \sim}^{\textstyle <} }
\newcommand{\vev}[1]{ \left\langle {#1} \right\rangle }
\newcommand{\bra}[1]{ \langle {#1} | }
\newcommand{\ket}[1]{ | {#1} \rangle }
\newcommand{\fb}{\,{\rm fb}^{-1}}
\newcommand{\ev}{{\rm eV}}
\newcommand{\kev}{{\rm keV}}
\newcommand{\Mev}{{\rm MeV}}
\newcommand{\gev}{{\rm GeV}}
\newcommand{\tev}{{\rm TeV}}
\newcommand{\mev}{{\rm MeV}}
\newcommand{\meV}{{\rm meV}}
\newcommand{\mnu}{\ensuremath{m_\nu}}
\newcommand{\nnu}{\ensuremath{n_\nu}}
\newcommand{\mlr}{\ensuremath{m_{lr}}}
\newcommand{\acc}{\ensuremath{{\cal A}}}
\newcommand{\mav}{MaVaNs}
\newcommand{\disc}[1]{{\bf #1}} 
\newcommand{\mh}{{m_h}}
\newcommand{\hb}{{\cal \bar H}}

\title{Leptophilic Dark Matter}
\author{Patrick J. Fox}
\affiliation{Theoretical Physics Department, Fermi National Accelerator Laboratory\\
 Batavia, IL 60510,USA}
\author{Erich Poppitz}
\affiliation{Department of Physics, University of Toronto\\
Toronto, ON M5S 1A7, Canada}
\preprint{\small FERMILAB-PUB-08-505-T}
\date{\today}
\begin{abstract}
We describe a simple model of Dark Matter, which explains the PAMELA/ATIC excesses while being consistent with all present constraints.  The DAMA annual modulation signal can also be explained for some values of the parameters.  The model  consists of a Dark Sector containing a weakly coupled broken $U(1)$ gauge symmetry, under which only the Dark Matter state and the leptons  are charged.  
\end{abstract}

\maketitle

\section{Introduction}

By now the existence of a large, non-baryonic contribution to the energy density of the universe---Dark Matter---is well established.  The exact nature of this new type of matter is the subject of much speculation.  It is searched for, in many experiments, both directly through its scatterings with standard model (SM) particles and indirectly through its annihilations to SM states.  We concentrate our attention in this letter on the results of several of these experiments,  PAMELA and ATIC, which search for DM indirectly through its annihilations to electrons/postirons and protons/anti-protons, and to a lesser extent DAMA and CDMS, which look  for DM directly through its scattering off atoms.

Recently PAMELA, a satellite based experiment, reported results for the flux ratio of protons to anti-protons and for the flux ratio of positrons to the sum of electrons and positrons.  In the proton/anti-proton channel they see no significant deviation~\cite{Adriani:2008zq} from the prediction of anti-proton production from the propagation of cosmic-rays through the galaxy.  In the electron/postiron channel there appears to be a significant excess~\cite{Adriani:2008zr} starting around energies of $10\,\gev$ and continuing to the highest bins at $100\,\gev$.  Both results are compatible with previous experiments but with higher precision.

The ATIC balloon experiment collaboration~\cite{atic} measured the total flux of electrons plus positrons out to energies of order $1\,\tev$.  There is an excess over what is expected from cosmic rays, peaked around $400$-$500\,\gev$.  This is in agreement with the measurement of another balloon experiment PPB-BETS~\cite{Torii:2008xu}, which also observes a peak around $\sim 500\,\gev$.

These excesses may be explained by astrophysical processes, for instance nearby pulsars may be a source for high energy positrons and electrons~\cite{Hooper:2008kg}, or they could be due to  annihilation of DM in our galactic neighbourhood.  Assuming the latter possibility,  the above results seem to indicate that the main annihilation is to electrons and positrons and not to hadronic final states.  
One way this  can happen is if the DM does not annihilate directly to the SM but instead first annihilates to a new state which in turn decays to SM states.  If this new state is lighter than the proton, the final state will only contain leptons \cite{Cholis:2008vb,Cholis:2008qq}. Thus,  the lack of hadronic final states is determined by the spectrum of new states  \cite{Finkbeiner:2008qu,ArkaniHamed:2008qp,ArkaniHamed:2008qn,Pospelov:2008jd}. 

Here, we consider instead the possibility that due to a symmetry the new states only have tree-level couplings to leptons but not to gauge bosons or quarks: leptophilic dark matter.  A model  similar to this, gauging $\mu-\tau$ number, and thus giving no possible DAMA signal, was briefly considered in~\cite{Cirelli:2008pk}, and lepton-friendly models in the context of supersymmetry, have been examined in the past~\cite{Baltz:2002we,Chen:2008dh}; here, we build a simple model and examine if it is possible to explain these excesses within the leptophilic framework.

We begin, in  Section~\ref{sec:themodel}, by describing the symmetry and the resulting model.  In Section~\ref{sec:constraints}, we discuss the existing constraints on the model to arrive at the viable region of parameter space.  In Section \ref{sec:pamela}, we explain how this region of parameter space is not only consistent with constraints, but may also explain the excesses discussed above.  Since the Dark Sector of our model only has couplings to leptons, CDMS, which vetoes on electromagnetic recoils, will have less sensitivity than DAMA, which records both nuclear and electromagnetic recoils.  In Section \ref{sec:directdetection}, we discuss whether leptophilic models can explain why DAMA observes a modulated signal but CDMS does not see any signal and the region of parameters where this is possible. In Section~\ref{sec:conclusions}, we conclude by recalling the main features of the model in the two interesting regions of parameter space. Finally, we note  that the coupling of the Dark Sector to neutrinos follows from the symmetries of our model  and point out the possibility of detection of neutrino flux from dark matter annihilations. 

\section{The Model}\label{sec:themodel}

We now describe the model:  we add to the SM a Dark Sector (DS) which contains a new Abelian gauge symmetry, $U(1)_{DS}$.  There is a Dirac fermion charged under this group that is also odd under a DS-parity (all SM fields are even under DS-parity).  This state will be the Dark Matter  (DM), in general there may be additional fermions charged under the $U(1)_{DS}$ but we ignore them here.  The gauge group is broken by a scalar Higgs field, or perhaps by technicolor-like dynamics (we will be agnostic about the precise mechanism) and the DS fermion has a vector-like mass.  The DS, for the case of scalar breaking, has the Lagrangian:
\be
\label{lds}
\mathcal{L}_{DS}=-{1 \over 4}  F_{\mu\nu}^{\prime^2} + \overline{\chi}\gamma^\mu D_\mu\chi + \left| D_\mu \phi\right|^2 - M_\chi \overline{\chi} \chi -V_{DS}(\phi) ~.
\ee
The coupling between the SM and the DS is through the new gauge boson $U$, with field strength denoted by $F^\prime$ in (\ref{lds}), thus some fields in the SM must be charged under  $U(1)_{DS}$.  We postulate that the  $U$ gauge boson is leptophilic and for anomaly cancellation require that  it couples with equal  and opposite charge to two generations of leptons. To
 allow SM Yukawa couplings,  the $U$-boson couplings to leptons are vectorlike; thus, the $U$-boson couples to neutrinos.

All that remains is to discuss the size of the couplings and masses in the problem.  First, we have the mass of the dark matter, $M_\chi$ and the $U$-boson, $M_U$. We also have the gauge couplings of the leptophilic gauge boson $U$ with the DM state $\chi$, $g_\chi$, and with the SM leptons, $g_l$. We will see that many of these parameters are tightly constrained by various experimental observations, making this model very predictive. 

In order to explain the PAMELA and ATIC excesses, the dark matter must have mass  larger than $\sim \mathcal{O}(700\,\gev)$.
Depending on the particle physics model, the parameters of the propagation model, the boost factor, and the dark matter distribution in the galaxy, the dark matter  may be also significantly heavier, e.g., in the few-$\tev$ range \cite{Cirelli:2008id}. However, given the uncertainties of these quantities, the mass can be close to the low value mentioned above---see the recent work \cite{Cholis:2008wq, Mardon:2009rc, Meade:2009rb} for a detailed model-independent analysis of the constraints and  uncertainties. Our interest here will be in the lower end of the allowed range, i.e.~dark matter mass $M_\chi \sim 700-800\, \gev$. 

 The annihilation cross section of DM into two $U$-bosons (we ignore the annihilation channel directly into two leptons, as in the parameter regime we are interested in this is small) is then: 
\be\label{eq:annihilation}
\langle\sigma_{ann}v\rangle= g_\chi^4\left(\frac{800\,\gev}{M_\chi}\right)^2 \times 31\, \mathrm{pb},
\ee
and the relic abundance can be explained with $g_\chi\sim 0.4$ and $M_\chi$ $\sim$ $700$-$800\,\gev$.  However, an annihilation cross-section of $\sim 1\,\mathrm{pb}$ yielding the correct relic abundance is too small to explain the PAMELA/ATIC excess; we will discuss the resolution in Section~\ref{sec:pamela}.  Before doing so, we will discuss constraints on the coupling of the $U$-boson to the SM leptons. 

\section{Constraints on $g_l$}\label{sec:constraints}

We have already described how the DM will freeze out with the correct relic abundance. However, without a coupling to the SM it may never get into equilibrium and certainly will lead to no observable signals.  The coupling of the $U$-boson to leptons will allow both of these to occur.  As already explained,  the $U$-boson has vectorlike couplings to two of the three SM generations; if there were a fourth generation \cite{Holdom:2006mr, Kribs:2007nz} this coupling could, in principle, be extended to include all generations.  

The size of the $U$-lepton coupling is strongly constrained by measurements of lepton magnetic-dipole moments and various leptonic cross sections \cite{Fayet}.
The contribution to a lepton anomalous magnetic dipole moment is given by:
\be
\Delta (g-2)_l \sim \frac{g_l^2}{4\pi^2}\frac{m_l^2}{M_U^2}
\ee
For the electron, muon, and tau, these are constrained to be smaller than $\sim 10^{-11}$, $\sim 10^{-9}$, and $\sim 10^{-2}$, respectively.  
Thus, the $U$-boson lepton couplings must obey:
\be
g_e\ltap 4\times 10^{-2}\frac{M_U}{\gev}, \ g_\mu\ltap 2\times 10^{-3}\frac{M_U}{\gev}, \ g_\tau \ltap 0.4 \frac{M_U}{\gev}~.
\ee
Furthermore, since the U-boson has a vectorlike coupling, it couples to neutrinos, allowing us to constrain it  from $\nu$-$e$ scattering at low $q^2$ \cite{Auerbach:2001wg}, yielding: 
\be
\label{gebound}
g_e \ltap 3\times 10^{-3} \frac{M_U}{\gev}~.
\ee
Finally, there are also constraints from $ee\rightarrow \gamma U$. At LEP, for couplings of order (\ref{gebound}) these are not significant.  $B$-factories, on the other hand, have the potential to place stronger bounds \cite{BBB}.  Using~\cite{:2008st} we find that for $M_U \le 7.8\,\gev$ the bound is $g_e\ltap 10^{-3}$, for particular values of $M_U$ this bound improves by a factor of $\sim 2$. 

From these constraints, we see that if the $U$-boson does not couple to the muon (hence it must couple to the electron and tau with opposite charge) we can avoid the strongest constraints from $g-2$, but the coupling $g_l$ is  appreciably smaller than $g_\chi$.  One might wonder how this can be explained?   We list several possibilities below:
\begin{itemize}
\item Since the group is  a $U(1)$ there is no technical reason why two different fields can not have wildly different charge.
\item Perhaps the DM state is a bound state of many unit charged objects \cite{Froggatt:2008uy}.
\item It is possible that the lightness of the leptons is due to a seesaw mechanism with some very heavy extra SM generations, that have unit charge under the extra $U(1)$.  If the SM leptons did not carry $U(1)$ charge but instead mixed with the heavy states through non-renormalisable operators then the smallness of the electron coupling would be due to the small mixing of the SM electron with the heavy state.
\item If the extra $U(1)$ is in a warped extra dimension setup, like that of Higgsless models~\cite{Csaki:2003zu}, then the lowest KK mode of the $U$-boson will have a wavefunction profile in the extra dimension such that it is suppressed at the IR brane.  If the leptons are confined to the IR brane and the DM is on the UV brane this may explain the large hierarchy in couplings.
\item Kinetic mixing, with coefficient $\kappa$, of the $U$ boson with another heavy gauge boson of mass $M$, which couples to a  lepton  current $J_\nu^{lept.}$ with couplings of order unity,  leads to  $U$-boson/lepton interactions of the form ${\kappa\over M^2} \partial_\nu F_{\mu\nu}^\prime J_\mu^{lept.} \sim {\kappa M_U^2 \over M^2} U^\nu J_\nu^{lept.}$, yielding sufficiently small couplings.
\end{itemize}
Whatever the reason for the smallness of $g_l$, if the DS is this simple, its couplings are well constrained by the observables described above.

Finally,  while at tree-level the $U$ gauge boson only couples to SM leptons and the DS, further couplings will be induced at the loop level. The most relevant is the  kinetic mixing   \cite{Holdom:1985ag} between the photon and $U$  through a  loop of SM leptons.  The mixing between the field strength of  $U$, $F^\prime_{\mu\nu}$, and the photon field strength, $F_{\mu\nu}$, is proportional to:
\be
\label{uphotonmixing}
\left[ \epsilon_{UV} + \frac{e g_e}{16\pi^2}\log \left(\frac{m_\tau}{m_e}\right)\right]  {F}^\prime_{\mu\nu} F^{\mu\nu}~,
\ee
where $\epsilon_{UV}$ denotes possible UV contributions to $U$-$\gamma$ mixing and the $\log$-enhanced contribution is the calculable IR contribution, written  under the assumption that the $U$-boson couples to $e$ and $\tau$. Without assuming any significant UV/IR cancellation, the $U$-boson  coupling to a charge-$q$ particle due to (\ref{uphotonmixing})  is then $g_q\sim 10^{-2} g_e q$---two orders of magnitude weaker than the coupling to leptons. This small coupling to quarks will not affect the branching ratio of $U$ to leptons, and thus the explanation of the PAMELA/ATIC excess, but it has implications for direct detection of DM, as we discuss below.
 
 In the simplest version of the model, there are two new states  in addition to the DM: $\phi$ and $U$.  In order that the abundances of light elements not be altered, the lifetime of these new states must be less than $\sim 1s$ such that they decay before BBN occurs.  The scalar is heavy and will quickly decay to SM leptons, the $U$ is light and has small couplings to SM leptons.  However, it is still far too short-lived to be a problem for BBN. The $U$-lifetime, of order   $ { 8 \pi \over M_U g_e^2 } \sim 10^{-17} s\; ({\gev \over M_U}) ({ 10^{-3} \over g_e})^2$, is also too short to significantly affect the energy loss of stars, even for $M_U$ as low as  $10$-$100\, \mev$.

\section{INDIRECT DETECTION}\label{sec:pamela}

Remarkably, even with these tight constraints on the $U$-boson couplings to the SM, it is still possible to explain the PAMELA and ATIC excesses.  These excesses are in electron and positron channels and not in hadronic channels \cite{Adriani:2008zq,Adriani:2008zr}. This is explained by the DM annihilating into $U$-bosons, which then decay into lepton pairs.  However, the annihilation cross section (\ref{eq:annihilation}) that gives the correct thermal abundance is not large enough to explain the observed rate (see~\cite{Nomura:2008ru, Nelson:2008hj, Harnik:2008uu} for  alternatives), but may be enhanced  \cite{sommerfeld} when the DM is slow moving, if there is a long-range attractive force between the annihilating states \cite{Hisano:2004ds,Hisano:2006nn,Cirelli:2007xd,ArkaniHamed:2008qn,MarchRussell:2008yu,Pospelov:2008jd,Cirelli:2008jk,Cirelli:2008pk,Fairbairn:2008fb,Bai:2008jt}.  The DM in our model is made of equal numbers (assuming no initial asymmetry) of positive and negative charged $\chi$, thus the exchange of $U$ is attractive for $\chi$-$\chi^c$.  For the attractive force to be sufficiently long range $M_U \ltap M_\chi g_\chi^2/4\pi \sim \mathcal{O}(10\,\gev)$. 
\begin{figure}
\begin{center}
\psfrag{muingev}[][][1.5]{$M_U\, [\gev]$}
\psfrag{mdmingev}[][][1.5]{$M_\chi\, [\gev]$}
\psfrag{0.1}[][][1.2]{$0.1$}
\psfrag{1}[][][1.2]{$1$}
\psfrag{10}[][][1.4]{$\!\!10$}
\psfrag{500}[][][1.4]{$500$}
\psfrag{600}[][][1.4]{$600$}
\psfrag{700}[][][1.4]{$700$}
\psfrag{800}[][][1.4]{$800$}
\psfrag{900}[][][1.4]{$900$}
\psfrag{1000}[][][1.4]{$1000$}
\includegraphics[width=3in]{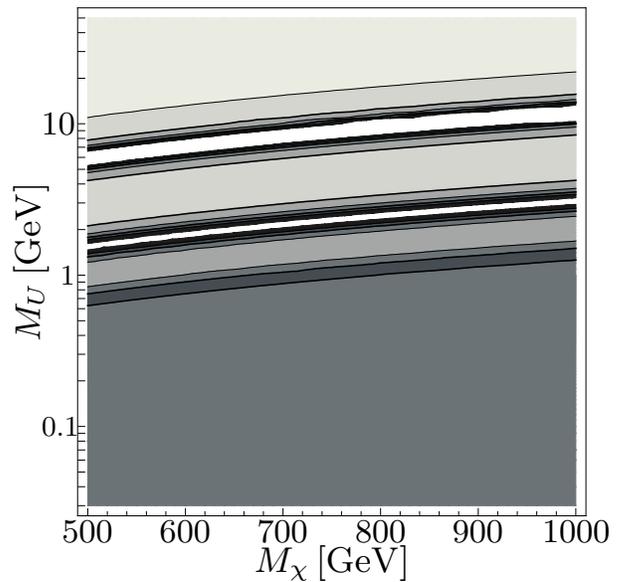}
\caption{Contour plot of the cross section boost factor as a function of the Dark Matter mass, $M_\chi$, and the $U$-boson mass, $M_U$.  The contours from light to dark grey correspond to enhancements of less than $10,\, 50,\,100,\, 150,\, 200$ while white corresponds to more than $200$, and we have taken $v=10^{-3}$ and $g_\chi=0.5$.}\label{fig:plot}
\end{center}
\end{figure}
Thus, in order for the Sommerfeld enhancement to work whilst still getting the correct thermal abundance of DM, there must be a hierarchy in the masses of the DM sector of at least an order of magnitude.    

In Figure~\ref{fig:plot}, we show the enhancement to the annihilation cross section due to the attractive force between the DM from the exchange of $U$-bosons; we have taken $v=10^{-3}$ and $g_\chi=0.5$.  Over most of the parameter space the boost factor is large ($\gtap 50$) but there are also regions where there are weakly bound resonances and the boost factor is far larger ($\gtap 1000$).  In addition to the boost factor due to the low-velocity enhancement there may be a boost factor due to an increase in the local abundance of DM which typically are order a few but may be as large as 13~\cite{Diemand:2006ik}.  For a DM mass of $\mathcal{O}(800\,\gev)$, that decays with a large branching fraction to $e$ and $\tau$, this gives sufficient enhancement to the cross section for reasonable choices of diffusion parameters~\cite{Cholis:2008hb}.  

Unlike the case of intermediate scalars \cite{Cholis:2008vb, Cholis:2008qq}, the injection spectrum of electrons in this case is not flat.  Instead, because the intermediate state is a vector, it is peaked to high and low values of energy.  When doing a full analysis of the propagation of the initial leptons to us this may slightly alter the best fit value of the dark matter mass.

Finally, we have a complete model of DM that can explain the electron-positron signals due to the fact that it only couples to electrons and taus, and their neutrinos.  The DM has mass around 800 GeV, there is another light state, $U$, of mass 1-10 GeV.  However, because the DM mainly couples to leptons it will be very hard to observe in direct detection experiments, which we discuss in the next section.  The final states of DM annihilation consist only of leptons, not because of the spectrum of the states involved~\cite{ArkaniHamed:2008qp,ArkaniHamed:2008qn,Nelson:2008hj,Nomura:2008ru} but rather because there is a symmetry forbidding anything else.  This allows for the force carrier to be heavier, the only constraint on its mass coming from the requirement of sufficient enhancement of its annihilation cross section.

\section{Direct Detection}\label{sec:directdetection}
  
If the DM couples only to leptons, almost all direct detection experiments will not be able to observe it, since they veto on leptonic recoils \cite{CDMS}.  The one exception is DAMA, which accepts all types of recoils and extracts the DM signal through its characteristic annual modulation.  One is then lead to wonder whether the PAMELA/ATIC excess is related to the DAMA-CDMS discrepancy.  Here we discuss the region of parameter space of our model that would lead to a DAMA signal.

The DAMA signal can be explained  if the DM-lepton elastic scattering cross section is of order $1\,\mathrm{pb}$ \cite{Bernabei:2007gr}. 
The $U$-mediated DM-electron cross section is: 
\bea
\label{damaxsection}
 \sigma^0_{DM-e}&&\equiv \frac{|\overline{\mathcal{M}}|^2}{16\pi M_\chi^2}
= \frac{g_\chi^2 g_e^2}{ \pi} \frac{m_e^2}{M_U^4} \nonumber \\
&&
= 0.5\, \mathrm{pb} \left(\frac{g_\chi}{0.4}\right)^{2} \left(\frac{g_e}{3\times10^{-5}} \right)^{2} \left(\frac{10\, \mathrm{MeV}}{M_U}\right)^{4}
\eea
where $\sigma^0_{DM-e}$ is the total cross section for scattering of non-relativistic dark matter off a stationary electron. 
Thus, for  $M_U = \mathcal{O}(10\,\mev)$, $g_\chi \sim 0.5$, $g_e\sim 10^{-5}$, consistent with the constraints of Section~\ref{sec:constraints} and the requirement of thermal abundance and positron signal, DAMA would have an observable signal \cite{Bernabei:2007gr}. 

To avoid a conflict with the lack of direct detection by CDMS,  the $U$-photon mixing parameter (\ref{uphotonmixing})  must be small enough to suppress the $U$-quark coupling  and, hence, the DM-nucleon cross section. 
The ratio of the DM-nucleon to the DM-electron  cross section scales as:
\be
\label{damavscdms}
 {\sigma^0_{DM-N} \over \sigma^0_{DM-e}} \sim \left({ g_q \over g_e } \right)^2 \left({ m_N \over m_e } \right)^2  \sim \left({ g_q \over g_e } \right)^2   \times 10^6~.
\ee
Now, CDMS \cite{CDMS} requires the  DM-nucleon cross section be less than $\sim 2\times 10^{-43}$ cm$^2$ for a $700$-$800\, \gev$ DM mass,
while the DM-electron cross section which allows for a DAMA signal, see eqn.~(\ref{damaxsection}) and \cite{Bernabei:2007gr}, is $10^{-36}$ cm$^2$, six orders of magnitude larger.
Thus, consistency with both experiments requies  $g_q \ltap 10^{-6} g_e$, implying a significant cancellation between an unspecified contribution from higher-scale physics, $\epsilon_{UV}$, and the infrared contribution to the $U$-photon mixing in (\ref{uphotonmixing}) (here, we will not address the question of how or whether this may naturally occur).
  
There are strong constraints coming from galactic dynamics~\cite{Ackerman:2008gi} on the strength of long-range DM-DM interactions.  Even for a light mediator, $M_U\sim 10\,\Mev$, the force is still sufficiently short range that there are not enough hard scatters to alter the momentum distribution of the DM halo.   
  
On the other hand, if we are to only explain the PAMELA/ATIC excesses, as  discussed in Section \ref{sec:pamela},  a much heavier $U$-boson of mass $M_U \sim 10\, \gev$  gives sufficient enhancement of the annihilation cross section. The bound  (\ref{gebound})  from low-energy measurements for this value of $M_U$ is  $g_e \ltap 10^{-2}$.  Taking tree-level  couplings of the $U$-boson of $g_\chi \sim 0.5$, $g_e \sim 10^{-4}$, while the one-loop coupling to quarks is $g_q \sim 10^{-6}$, as expected from the IR contribution in (\ref{uphotonmixing}) alone, 
we find from (\ref{damaxsection}) and (\ref{damavscdms}) a DM-electron cross-section $\sigma_{DM-e} \sim 10^{-47}$cm$^2$, while the DM-nucleon cross section is $\sigma_{DM-N} \sim 10^{-45}$cm$^2$, 
consistent with the current CDMS bounds and within reach of planned direct detection experiments.

Thus, in our model, only the DM has an order one  coupling  to the $U$-boson.  Note that if, due to cancellation with physics in the UV, the effective $U$-$\gamma$ mixing were tiny then the strong constraint from CDMS would go away and the dominant constraint on the size of $g_e$ would be due to $\nu$-$e$ scattering, i.e. $g_e\ltap 10^{-2}$.
In \cite{ArkaniHamed:2008qn}, the $U$-boson does not couple to neutrinos and this strong constraint is missing.  But, unlike here, in \cite{ArkaniHamed:2008qn}  the $U$-boson couples directly to quarks and then there is a strong constraint from the lack of a signal at CDMS, requiring an equally small coupling of $U$ to quarks, $10^{-5}$.  This can be avoided in  \cite{ArkaniHamed:2008qn} if the DM only scatters inelastically; for us the scattering is elastic but mainly off electrons.  Since we have a symmetry forbidding DM-annihilation into hadrons, rather than kinematics, we are able to have a larger mediator mass, allowing us to avoid the potential constraints from diffuse gamma-ray backgrounds~\cite{Kamionkowski:2008gj}.  Assuming that the DM halo profile smoothly extrapolates to the inner region of the galaxy, it is expected that the galactic center and galactic ridge will have a significantly increased dark matter density and may be significant sources of photons~\cite{Bertone:2008xr, Bergstrom:2008ag}.  However, there is considerable uncertainty in this extrapolation of dark matter density and velocity profiles.  In addition a cascade decay of the DM softens the spectrum of produced photons, relative to that of direct decay.  These effects have the potential to evade the constraints coming from the lack of observation of gamma rays from the inner few 100 pc of the galaxy~\cite{Meade:2009rb,Mardon:2009rc}.

\section{Conclusions}\label{sec:conclusions}

We have  constructed a model to explain the results reported by the PAMELA, ATIC, and PPB-BETS experiments, namely several leptonic excesses and at the same time the seeming lack of anti-proton excesses.  We have taken an extreme intepretation of their results, that the DM can not annihilate, at tree-level, into hadrons but only into leptons.  Rather than enforce this by a hierarchy in the DS, with the DM decaying to a very light mediator, we have instead enforced this difference by means of a symmetry.  We gauged a flavor dependent lepton number symmetry under which the DM, a Dirac fermion, is also charged.  This results in the DM annihilating into electrons and either muons or taus (here we considered the case of decays into $e$ and $\tau$). 

New couplings to electrons are tightly constrained by various measurements: anomalous magnetic moments, LEP and B-physics searches, and low energy $\nu$-$e$ scattering.  However, we showed that it is possible to satisfy all these constraints while explaining the leptonic excesses.  Unless there is cancellation with UV physics, loop-level couplings of the DM to hadrons will be induced, leading to further constraints coming from the lack of detection at CDMS. We described a region of parameter space where these constraints are also satisfied and the explanation of the leptonic excesses is maintained.  Finally, we also pointed out that it is possible, if the hadronic coupling is tiny, that CDMS would veto the leptonic scatters and only DAMA would have sensitivity to directly detect the DM. We described a particular corner of parameter space  where this is possible.

In addition to annihilating to charged leptons, the leptophilic DM also annihilates to neutrinos, a distinction from many other models with light mediators.  Should the DM be captured in the sun, an open question given it only has sizeable couplings to leptons, is whether it is possible for experiments such as ICECUBE \cite{Klein:2008px}  to search for neutrinos from DM annihilations in the sun's interior.    Since the leptons now carry a charge under the new $U(1)$ it would be interesting to see if this charge can explain the pattern of neutrino mixing angles.  Collider searches for dark matter in this model will be difficult, due to the tiny coupling to leptons and quarks, unless there are further modifications to this very minimal model.  For instance, a UV completion of the model may  introduce further couplings between the DS and SM, suppressed by a higher scale, as in the ``hidden valley" framework~\cite{Strassler:2006im}; if such couplings are present, lepton jets  \cite{ArkaniHamed:2008qp} may be observed in colliders. 

\section{Acknowledgements}
We thank Yang Bai, Rikard Enberg, Yuri Kolomensky, Bob McElrath, Vivek Sharma, Neal Weiner, and especially Kathryn Zurek for discussions. EP acknowledges support by the National Science and Engineering Research Council of Canada (NSERC).  Fermilab is operated by Fermi Research Alliance, LLC, under Contract DE-AC02-07CH11359 with the United States 
Department of Energy.


\end{document}